\documentclass[conference]{IEEEtran}
\usepackage{cite}
\ifCLASSINFOpdf
\usepackage[pdftex]{graphicx}
\graphicspath{./figures}
\DeclareGraphicsExtensions{.pdf,.jpeg,.png}
\else
\fi
\usepackage[cmex10]{amsmath}
\usepackage{array}
\usepackage{url}
\usepackage[utf8]{inputenc}  
\usepackage{citesort}
\usepackage[T1]{fontenc} 
\usepackage{mathtools}
\usepackage{amsfonts}
\usepackage{amssymb}

\DeclareUnicodeCharacter{00A0}{~}

\def \({\left(}
\def \){\right)}
\def \[{\left[}
\def \]{\right]}
\newcommand{\tbf}[1]{{\textbf{#1}}}

\newcommand{\txt}[1]{\text{#1}}
\newcommand{\defeq}{\vcentcolon=}

\newcommand{\bF}{{\textbf {F}}}

\newcommand{\bX}{{\textbf {X}}}
\newcommand{\bx}{{\textbf {x}}}

\newcommand{\by}{{\textbf {y}}}
\newcommand{\bz}{{\textbf {z}}}

\newcommand{\bs}{{\textbf {s}}}

\newcommand{\be}{\begin{equation}}
\newcommand{\ee}{\end{equation}}
\newcommand{\bea}{\begin{eqnarray}}
\newcommand{\eea}{\end{eqnarray}}

\newtheorem{theorem}{Theorem}[section]
\newtheorem{lemma}[theorem]{\textbf{Lemma}}
\newtheorem{thm}[theorem]{\textbf{Theorem}}

\newtheorem{corollary}[theorem]{\textbf{Corollary}}
\newtheorem{definition}[theorem]{\textbf{Definition}}

\begin{document}
\title{Proof of Threshold Saturation for \\Spatially Coupled Sparse Superposition Codes}
\author{\IEEEauthorblockN{Jean Barbier, \emph{Member IEEE}, Mohamad Dia and Nicolas Macris, \emph{Member IEEE}.}\\
\IEEEauthorblockA{Laboratoire de Th\'eorie des Communications, Ecole Polytechnique F\'ed\'erale de Lausanne.\\ \{jean.barbier, mohamad.dia, nicolas.macris\}@epfl.ch}}
\markboth{Threshold Saturation for Spatially Coupled Sparse Superposition Codes}%
{Threshold Saturation for Spatially Coupled Sparse Superposition Codes}
\maketitle
\IEEEpeerreviewmaketitle
\begin{abstract}
Recently, a new class of codes, called sparse superposition or sparse regression codes, has been proposed 
for communication over the AWGN channel. It has been 
proven that they achieve capacity using power allocation and 
various forms of iterative decoding. Empirical evidence has 
also strongly suggested that the codes achieve capacity when 
spatial coupling and approximate message passing decoding are used, without need of power allocation. 
In this note we prove that state evolution (which tracks 
message passing) indeed 
saturates the potential threshold of the underlying code ensemble, which approaches in a proper limit the optimal threshold. Our proof 
uses ideas developed in the theory of low-density parity-check codes and compressive sensing. 
\end{abstract}
%
%
\section{Introduction}
Sparse superposition (SS) codes 
introduced by Barron and Joseph for reliable communication over the additive white Gaussian noise channel (AWGNC)
have been proven to approach
capacity using power allocation and various efficient decoders \cite{barron2010sparse,JosephB14,barron2012high}. 
An approximate message passing (AMP) decoder was introduced in 
\cite{barbier2014replica}, and the recent analysis \cite{rush2015capacity} proves that this allows to reach 
capacity with the help of power allocation. 
Spatially coupled SS codes were introduced in \cite{barbierSchulkeKrzakala,BarbierK15} and empirically shown 
to reach capacity under AMP {\it without any need for power allocation}. 
The empirical evidence shows that 
spatially coupled SS codes perform better than power allocated ones in the sense that they approach 
capacity faster in a proper limit \cite{BarbierK15}. Given this evidence, it is of interest to develop a rigorous theory for such coding constructions. 

It is natural to address two conjectures. First, that
spatially coupled SS codes allow to reach the so-called 
\emph{potential threshold} of state evolution (SE).
Second, that SE correctly tracks the AMP decoder. As we will argue, the potential threshold tends to capacity in a proper limit, so this would prove that the codes are capacity achieving. 

The purpose of this note is to settle the first conjecture. 
We prove that for a general ensemble of spatially coupled 
coding matrices, the AMP threshold  attains (in a suitable limit) the potential threshold of SE. This phenomenon is called {\it threshold saturation}.
The precise statements of our main results are 
Theorem~\ref{th:mainTheorem} and Corollary~\ref{cor:maincorollary}. 

Threshold saturation was first established 
in the context of spatially coupled Low-Density Parity-Check codes for general binary input memoryless symmetric channels 
in \cite{Kudekar-Urbanke-Richardson-2013}, and is recognized as the basic mechanism underpinning the excellent performance of 
such codes \cite{Zigangirov-Costello-2010}. It is interesting that essentially the same phenomenon can be established for 
a coding system operating on a channel with {\it continuous inputs}. 
This result
is a stepping-stone towards establishing that spatially coupled SS codes achieve capacity on the AWGNC under 
AMP decoding. The remaining analysis to settle the second conjecture would require extending 
the one given for
compressive sensing \cite{Montanari-Javanmard} to signals with correlated components in a spatially coupled system, as already 
done for the power allocated system \cite{rush2015capacity}.

To establish threshold saturation, we use the \emph{potential method} along the lines of \cite{YedlaJian12,PfisterMacrisBMS} developed for LDPC and 
LDGM codes. Note that a similar (but different) potential to the one used here has been introduced in the context of scalar
compressive sensing \cite{BayatiMontanari10,6887298}. It is interesting that the potential method 
goes through for the present system involving a dense coding matrix and a fairly wide class of spatial couplings \cite{KrzakalaMezard12}.

Coding constructions of the underlying and coupled ensembles are
described in Sec.~\ref{sec:codeens}. Sec.~\ref{sec:stateandpot} reviews SE and potential formulations adapted to the present context. The AMP thresholds of underlying and coupled ensembles as well as the potential threshold are given precise definitions. 
Sec.~\ref{sec:propCoupledSyst} introduces a notion of degradation and settles monotonicity properties of SE.
The essential steps for the proof of threshold saturation are presented in Sec.~\ref{sec:proofsketch}.

%
\section{Code ensembles}\label{sec:codeens}
%
%
We first define the underlying and spatially coupled ensembles of SS codes for transmission over an AWGNC.
We will often use the shorter notations $[a_1 : a_n]$ and $\{b_1 : b_n\}$  instead of $[a_1, \dots, a_n]$ and $\{b_1, \dots ,  b_n\}$ for $n$-tuples.
\subsection{The underlying ensemble}
In the framework of SS codes, the \emph{information word} $\bs = [\bs_1 : \bs_L]$ is a vector made 
of $L$ \emph{sections}. Each section $\bs_l$, $l\in\{1:L\}$,  is a $B$-dimensional vector with one component
equal to $1$ and $B-1$ components equal to $0$. For example if $(B=3,L=4)$, a valid information word is $\tbf s = [001,010,100,010]$. We call $B$ the \emph{section size} (or alphabet size) and set $N=LB$. 
A \emph{codeword} $\bF\bs \in \mathbb{R}^{M}$ is generated from a fixed \emph{coding matrix} $\bF\in \mathbb{R}^{M \times N}$. 
We consider random codes generated by $\bF$ drawn from the ensemble of random matrices with i.i.d real Gaussian entries
with distribution $\mathcal{N}(0, 1/L)$.
The cardinality of the code is $B^L$, the block length is $M$, and the (design) rate is $R=L\log_2(B)/M = N\log_2(B)/(M B)$. The code is thus specified by the basic parameters $(M, R, B)$.

Codewords are transmitted through an AWGNC, i.e., 
the received signal is $\by= \bF\bs +\boldsymbol\xi$ where $\xi_\mu \sim\mathcal{N}(0, \sigma^2) \ \forall \ \mu$.
Power is normalized thanks to the $1/L$ variance of the entries of $\bF$, so that the signal-to-noise ratio is ${\rm snr = 1/\sigma^2}$. 

The decoding task is to infer the information word $\bs$ from channel observations $\by$. This is obviously very similar in spirit to compressive sensing, where one wants to infer a sparse signal from a certain number of measurements. For this reason, our language and techniques are sometimes borrowed from compressive sensing. In particular, the code rate can be linked to a 
``measurement rate'' $\alpha \defeq M/N = \log_2(B)/ (BR)$.
\subsection{The spatially coupled ensemble}
\begin{figure}[!t]
\centering
\includegraphics[width=.45\textwidth]{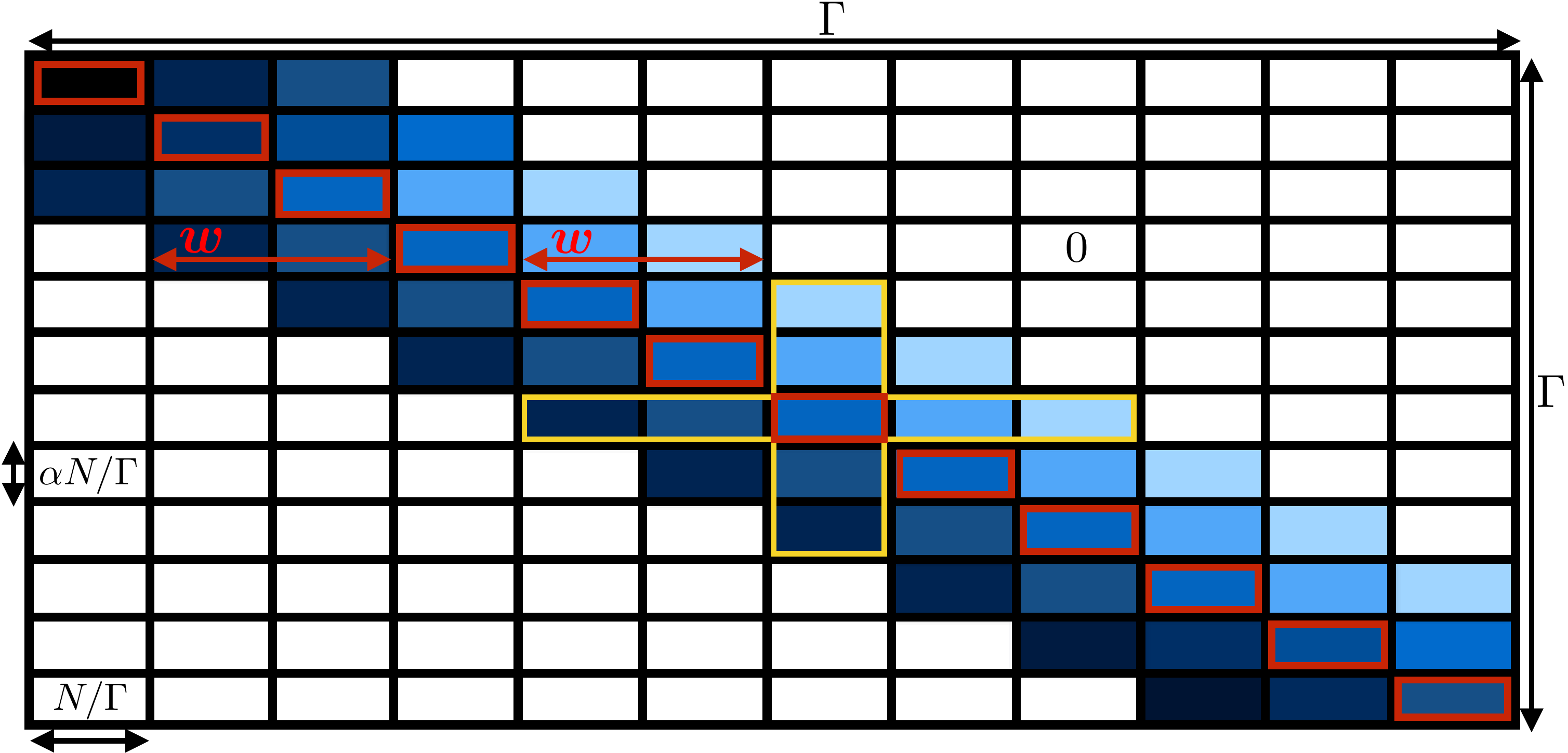}
\caption{A spatially coupled coding matrix $\in\mathbb{R}^{M\times N}$ made of $\Gamma\times \Gamma$ blocks indexed by $(r,c)$, each with $N/\Gamma$ columns and $M/\Gamma =\alpha N/\Gamma$ rows. The i.i.d elements in block $(r,c)$ have distribution $\mathcal{N}(0,J_{r,c}/L)$. Away from the boundaries, in addition to the diagonal (in red), there are $w$ forward and backward coupling blocks. In this example, the design function $g_w$ enforces a stronger backward coupling. Blocks are darker at the boundaries because the variances are larger so as  to enforce the \emph{variance normalization} $\sum_{c=1}^\Gamma J_{r,c}/\Gamma = 1 \ \forall \ r$. The yellow shape emphasizes \emph{variance symmetry} $\sum_{r=1}^\Gamma J_{r,k}/\Gamma=\sum_{c=1}^\Gamma J_{k,c}/\Gamma=1$ verified if $k\in \{2w+1:\Gamma-2w\}$.}
\label{fig:opSpCoupling}
\end{figure}
The construction has similarities with \cite{CaltagironeZ14}. We consider spatially coupled codes based on coding 
matrices in $\mathbb{R}^{M\times N}$ as described in details in Fig.~\ref{fig:opSpCoupling}. This ensemble of matrices 
is parametrized by $(M,R,B,\Gamma,w,g_w)$, where $w$ is the \emph{coupling window} and $g_w$ is the \emph{design function}. 
This is any function verifying $g_w(x) = 0$ if $|x|>1$ and $g_w(x)\ge g_0>0$ else, which 
is Lipschitz continuous on its support with Lipschitz constant $g_*$. The constants $g_0$, $g_*$ are independent of $w$. Moreover, we impose the normalization
 $\sum_{k=-w}^{w} g_w(k/w)/(2w+1) = 1$.

From $g_w$, we construct the \emph{variances} of the blocks: the i.i.d entries inside the block $(r,c)$ are distributed as $\mathcal{N}(0,J_{r,c}/L)$, where $J_{r,c} \defeq \gamma_r \Gamma  g_w((r-c)/w)/(2w+1)$. Here $\gamma_r$ enforces
the \emph{variance normalization} $\sum_{c=1}^{\Gamma}J_{r,c}/\Gamma = 1 \ \forall \ r$. 
%
%
This implies (by the law of large numbers) that $[\bF \bs]_\mu^2 \to 1 \ \forall \ \mu$ as $L\to\infty$, i.e. the asymptotic power $\lim_{L\to \infty}\sum_{\mu=1}^M [\bF \bs]_\mu^2/M = 1$, and the ${\rm snr} = 1/\sigma^2$ is homogeneous.

The spatial coupling induces a natural decomposition of the signal into $\Gamma$ \emph{blocks} (associated with the block-columns of the matrix), each made of $L/\Gamma$ sections. One key element of spatially coupled codes is the \emph{seed} introduced at the boundaries. 
We assume the sections in the first and 
last $4w$ blocks of the information word ${\tbf s}$ to be known by the decoder. This boundary 
condition can be interpreted as perfect side measurements that propagate inward and boost the performance. 
The seed induces a rate loss in the \emph{effective rate} of the code, $R_{\text{eff}} = R (1-\frac{8w}{\Gamma})$.
However, this loss vanishes as $\Gamma \rightarrow \infty$.
%
%
\section{State evolution and potential formulation}\label{sec:stateandpot}
We now give the {\it state evolution} associated to the underlying and spatially coupled ensembles,  which is conjectured to track the performance of the AMP decoder. We then define an appropriate {\it potential function} for each ensemble. 
\subsection{State evolution}
The goal is to iteratively compute the 
mean square error (MSE) $\tilde E^{(t)} =\frac{1}{L}\mathbb{E}_{\bs, \by}[\sum_{j=1}^N
(\hat{s}_j^{(t)} - s_j)^2]$ of the AMP estimates $\{\hat{s}_j^{(t)}\}$ at 
iteration $t\in \mathbb{N}$. We need a few definitions to express the iterations. Let 
$\Sigma(E) \defeq \sqrt{R(\sigma^2 +E)}$.
We define a {\it denoiser} $f_i(\Sigma(E))$ as 
the minimum mean square error (MMSE) estimator of the $i$-th component of a section sent through an {\it effective} 
AWGNC with a noise $\mathcal{N}(0, \Sigma(E)^2/\log_2(B))$, when the input signal is uniformly distributed. 
Explicitly, if $\bz = [z_1 : z_B]$ denote 
$B$ independent standardized Gaussian random variables (with zero mean and unit variance), and 
$p_0(\bx) = \frac{1}{B}\sum_{i=1}^B\delta_{x_i, 1}\prod_{k\neq i}^B\delta_{x_k, 0}$, we set for $i=1 : B$
\begin{align}
  & f_i(\Sigma(E)) \defeq 
 \frac{\sum_{\bx} e^{-\frac{\|\bx - (\bs +\bz \Sigma(E)/\sqrt{\log_2(B)})\|_2^2}{2\Sigma(E)^2/\log_2(B)}} p_0(\bx) x_i}
 {\sum_{\bx} e^{-\frac{\|\bx - (\bs +\bz \Sigma(E)/\sqrt{\log_2(B)}) \|_2^2}{2\Sigma(E)^2/\log_2(B)}} p_0(\bx)}
\nonumber \\ &
=
\Bigg[1+\sum_{k\neq i}^B 
e^{(s_k- s_i)\log_2(B)/\Sigma(E)^2 + (z_k - z_i)\sqrt{\log_2(B)}/\Sigma(E)}\Bigg]^{-1}.
\nonumber
\end{align}
Note that the denoiser also depends on $\bs, \bz$. 
Furthermore, we define the {\it SE operator of the underlying system} as
\begin{align}\label{equ:stateevopunderlying}
 T_\text{u}(E) \defeq \mathbb{E}_{\bs, \bz}\biggl[\sum_{i=1}^B \big(f_i(\Sigma(E)) - s_i\big)^2\biggr].
\end{align}
This is nothing else than the MSE associated to the MMSE estimator of the effective AWGNC with noise $\mathcal{N}(0, \Sigma(E)^2/\log_2(B))$. The SE iterations for the underlying system's MSE can now be expressed as 
\begin{align}\label{equ:stateevolutionunderlying}
\tilde E^{(t+1)} = T_\text{u}(\tilde E^{(t)}), \qquad t\geq 0.
\end{align}
To track the performance of the AMP decoder the iterations are initialized with $\tilde E^{(0)}=1$. The monotonicity and boundedness of the iterations of SE, discussed in Sec.~\ref{sec:propCoupledSyst}, ensure that actually all initial conditions reach a fixed point.

\begin{definition}[MSE Floor] The MSE floor $E_0$ is defined as the fixed point reached from the initial condition $\tilde E^{(0)} = 0$. In other words $E_0 = T_\text{u}^{(\infty)}(0)$. 
\end{definition}
%
%
\begin{definition}[Bassin of attraction]
The basin of attraction of the MSE floor $E_0$ is 
$\mathcal{V}_0 \defeq \big\{ E \ \! |\ \! T_\text{u}^{(\infty )}(E) = E_0 \big\}$.
\end{definition}

\begin{definition} [Threshold of underlying ensemble]
The AMP threshold is $R_{\text{u}} \defeq {\rm sup}\{R>0\ \! |\ \! T_{\text{u}}^{(\infty)}(1) = E_0\}$. 
\end{definition}

It can be shown that for the present system $T_{\text{u}}^{(\infty)}(0)$ and $T_{\text{u}}^{(\infty)}(1)$
are the only two possible fixed points. For $R<R_{\text{u}}$, there is only one fixed point, namely the ``good'' one
$T_{\text{u}}^{(\infty)}(0)$, 
and for large section size $B$ the MSE floor and section error rate are small. Instead if $R>R_{\text{u}}$, the decoder is blocked 
by the ``bad'' fixed point $T_{\text{u}}^{(\infty)}(1)\neq E_0$ and AMP cannot decode.

We now turn our attention to SE for the spatially coupled system. The perfomance of AMP is described by
an {\it MSE profile} $\{\tilde E_c \ \! |\ \! c= 1: \Gamma\}$ along the ``spatial dimension''. Since we assume that the boundary sections are known to the decoder, we enforce the \emph{pinning condition} $\tilde E_c =0$ for $c\in\{1: 4w\}\cup\{\Gamma-4w+1 : \Gamma\}$.
For $c$ not in this set, by definition $\tilde E_c \defeq \frac{\Gamma}{L} \sum_{l\in c} \mathbb{E}_{\bs, \by}[\|\hat{\tbf s}_l - {\tbf s}_l\|_2^2]$, where the sum $l\in c$ is over
the set of indices of the $L/\Gamma$ sections composing the $c$-th block of $\bs$. 
It is more convenient to 
work with a \emph{smoothed error profile} (referred as the error profile) ${\tbf E}= \{E_r \ \! | \ \! r=1 : \Gamma\}$, 
$E_r \defeq \frac{1}{\Gamma} \sum_{c=1}^{\Gamma} J_{r,c} \tilde E_c$.
Indeed, this change of variables makes the problem mathematically more tractable for spatially coupled codes.
The pinning condition becomes $E_r=0$ for $r\in\mathcal{A}\defeq\{1: 3w\}\cup\{\Gamma-3w+1: \Gamma\}$. 

We define an {\it effective noise} for block $c\in\{1: \Gamma$\},
\begin{equation}\label{equ:effectivenoisecoupled}
 \Sigma_{c}({\tbf E} )\defeq \biggl[\frac{1}{\Gamma}\sum_{r=1}^{\Gamma} \frac{J_{r,c}}{R(\sigma^2 +E_r)}\biggr]^{-1/2}
\end{equation}
and the {\it coupled SE operator}
\begin{equation}\label{equ:stateevopcoupled}
[T_{\rm c}({\tbf E})]_r = \frac{1}{\Gamma}
 \sum_{c=1}^{\Gamma}J_{r, c}\mathbb{E}_{\bs, \bz}\biggl[\sum_{i=1}^{B}\big(f_i(\Sigma_{c}({\tbf E} )) - s_i\big)^2\biggr].
\end{equation}
$T_{\rm c}({\tbf E})$ is vector valued and here we have written its $r$-th component. The SE iterations then read for $r\notin \mathcal{A}$
\begin{equation}\label{equ:statevolutioncoupled}
E_r^{(t+1)} = [T_{\rm c}({\tbf E}^{(t)})]_r, \qquad t\geq 0.
\end{equation}
For $r\in \mathcal{A}$, the pinning condition
$E_r^{(t)} =0$ is enforced at all times. SE is initialized with $E_r^{(0)}=1$ for $r\notin \mathcal{A}$.   

Let ${\tbf E}_0 \defeq [E_r=E_0, r=1: \Gamma]$ be the \emph{MSE floor profile}. 

\begin{definition} [Threshold of coupled ensemble]\label{def:AMPcoupled}
The AMP threshold of the coupled system is defined as $R_{\text c} \defeq {\liminf}_{\Gamma, w\to \infty} {\rm sup}\{R>0\ \! |\ \! T_{\text{c}}^{(\infty)}(\boldsymbol{1}) \prec \tbf E_0\}$
where $\boldsymbol{1}$ is the all ones vector. Here the ${\liminf}_{\Gamma, w \to \infty}$ is taken 
along sequences where {\it first} $\Gamma \to \infty$ and {\it then} $w\to\infty$. 
\end{definition}

%
%
\subsection{Potential formulation}\label{subsec:potentials}
The fixed point equations associated to SE iterations can be reformulated as stationary point equations for 
suitable {\it potential functions}. These are in general not unique. However, the ``correct'' guess 
(i.e. the one that allows to prove full threshold saturation) can be derived by the replica method of statistical physics \cite{barbier2014replica} .

The potential of the underlying ensemble 
 is $F_{\text{u}}( E) \defeq U_{\text{u}}(E) - S_{\text{u}}( \Sigma( E))$, with
\begin{align}
\begin{cases}
U_{\text{u}}(E) \defeq \frac{1}{2R} \log_2\Big((\sigma^2 +  E) e^{-\frac{ E}{\sigma^2 +  E}}\Big), \\
S_{\text{u}}( \Sigma( {E})) \defeq \mathbb{E}_\bz\bigr[\log_B\big(1+\sum_{i=2}^B e_{i}(\Sigma( {E}))\big)\bigl],
\end{cases}
\end{align}
where $e_{i}(x) \defeq \exp\big((z_i-z_1)\sqrt{\log_2(B)}/x-\log_2(B)/x^2 \big)$. 

The potential of the spatially coupled ensemble is $F_{\text{c}}({\tbf E}) = U_{\text{c}}({\tbf E}) - S_{\text{c}}({\tbf E})$ 
where 
\begin{align}
\begin{cases}
U_{\text{c}}(\tbf{E}) \defeq \sum_{r=1}^{\Gamma} U_{\text{u}}( E_r), \\
S_{\text{c}}(\tbf{E}) \defeq\sum_{c=1}^{\Gamma} S_{\text{u}}( \Sigma_c({\tbf E})).
\end{cases}
\end{align}

Let us pause for an instant and give the 
statistical physics interpretation of these formulas. The posterior distribution 
$p(\bs\vert \by) = \exp(-\|\by - \bF\bs\|_2^2/2\sigma^2) p_0(\bs) /Z$ ($Z$ the normalizing factor) can be interpreted  as 
the Gibbs distribution of a disordered spin system ($\bs$ being the annealed ``spin'' degrees of freedom, $\by$ and $\bF$ the quenched ``disorder''). 
The potential functions $F$ are ``Bethe free energies'' averaged over the disorder. 
They are equal to ``energy'' terms $U$ minus ``entropy'' terms $S$. One can prove that both 
terms are increasing in $\Sigma$. This is ``physically'' natural if the effective channel noise $\Sigma$ is 
interpreted as a kind of effective ``temperature''. 
%
%
\begin{definition}[Free energy gap]
The free energy gap is $\Delta F_{\text{u}} \defeq {\rm inf} _{E \notin \mathcal{V}_0 } (F_{\text{u}}( E) - F_{\text{u}}(E_0))$, with the convention that the infimum over the empty set is $\infty$  (this happens for $R < R_{\text{u}}$).
\end{definition}
\begin{definition}[Potential threshold]
\label{def:potThresh}
The potential threshold is $R_{\rm pot} \defeq {\rm sup}\{R>0\ \! |\ \! \Delta F_{\text{u}} > 0\}$. 
\end{definition}

The connection between these potentials and SE is given by the following Lemma.
\begin{lemma}\label{lemma:fixedpointSE_extPot}
If $\mathring{E}$ is a fixed point of \eqref{equ:stateevolutionunderlying}, i.e., 
$T_{\text u}(\mathring E) = \mathring E$,  then $[\frac{\partial F_{\text{u}}}{\partial E}]_{\mathring E} =0$. Similarly for the coupled system,
if $\mathring{{\tbf E}}$ is a fixed point of \eqref{equ:statevolutioncoupled} then $[\frac{\partial F_{\text{c}}}{\partial E_r}]_{\mathring{{\tbf E}}} = 0 \ \forall \ r\in \{3w+1 : \Gamma-3w\}$.
\end{lemma}

The proof is technical and we skip it here. Let us just indicate that it proceeds by computing the derivatives of the 
potentials, uses Gaussian integration by parts formulas and channel symmetry.
Similar results can be found in \cite{YedlaJian12,PfisterMacrisBMS}.

It is noteworthy that what is called a ``Bethe entropy'' in the statistical physics literature, has an information theoretic interpretation, and is actually a Shannon conditionnal entropy for 
an effective channel.
Let $\bX$ be a $B$-dimensional random vector with distribution $p_0(\bx) = \frac{1}{B}\sum_{i=1}^B\delta_{x_i, 1}\prod_{k\neq i}^B\delta_{x_k, 0}$. Take the 
output ${\tbf Y}$ of an AWGNC with i.i.d noise $\mathcal{N}(0, \Sigma^2/\log_2(B))$ for each component, when the input is $\bX$. Then it is an exercise to check that 
$H(\bX \vert {\tbf Y}) = S_{\text{u}}(\Sigma) \log_2(B)$. This identification clearly shows that $S_{\text{u}}(\Sigma)$ must be an increasing function 
of $\Sigma$. Also, it allows to give information theoretic expressions for the 
potential functions. We note that such expressions have already been written down for {\it scalar} compressive sensing \cite{BayatiMontanari10,6887298}. 

We also point out that there is another way to derive potential functions by integrating out the SE fixed point equations after premultiplication by an ``integrating factor''. When the correct ``integrating factor'' is used, one recovers the information theoretic expressions of the potential functions. This method is discussed in \cite{6887298} for a wide range of problems including scalar compressive sensing, and one can check
it extends to the present setting. A key point for this method is the well known relation between mutual information (or conditional entropy) and 
MMSE \cite{guo2011estimation}. Here this relation takes the form
$\frac{d H(\bX| {\tbf Y})}{d (\Sigma^{-2})} = - \frac{1}{2} {\rm mmse}(\Sigma)$ (here ${\rm mmse} = T_{\rm u}$ where $T_{\rm u}$ is the r.h.s of
\eqref{equ:stateevopunderlying} viewed as a function of $\Sigma$).

\subsection{Large $B$ analysis and connection with Shannon's capacity}
\label{sec:larg_B}

Let us now emphasize the connection 
between the potential threshold $R_{\rm pot}$ and Shannon's capacity $C= \frac{1}{2}\log_2(1+{\rm snr})$. 
The large section size limit of the potential of the underlying system becomes \cite{phdBarbier}
\begin{align}
\lim_{B\to\infty} F_{\text{u}}(E) = U_{\text{u}}(E) - {\rm max}\Big(0,1 - \frac{1}{2\ln(2)\Sigma(E)^2}\Big),
\end{align}
(where we recall $\Sigma(E)^2 \defeq R(\sigma^{2} +E)$). 
The analysis of this function of $E\in[0,1]$ shows the following. For $R<[(1+\sigma^2)2\ln(2)]^{-1}$ there is a unique minimum at $E=0$.
For $[(1+\sigma^2)2\ln(2)]^{-1} < R < C$, $E=0$ is the global minimum but there exists a local minimum at $E=1$. When $R>C$ the global minimum is 
at $E=1$ and $E=0$ is a local minimum. Therefore we can identify $\lim_{B\to \infty} R_{\rm pot} = C$ and $\lim_{B\to \infty} R_{\rm u} = [(1+\sigma^2)2\ln(2)]^{-1}$.

Let us also point out that these are static properties of the system which are independent of the decoding algorithms. In a sense they confirm in an alternative way that the codes must achieve capacity under optimal decoding \cite{leastsquareBarron}.

%
\section{Properties of the coupled system}\label{sec:propCoupledSyst}
Monotonicity properties of the SE operators $T_{\rm u}$ and $T_{\rm c}$ are key elements in the analysis. We start by defining a suitable notion of {\it degradation}. 
\begin{definition}[Degradation]
A profile ${\tbf{E}}$ is degraded (resp. strictly degraded) with respect to another one ${\tbf{G}}$, 
denoted as ${\tbf{E}} \succeq {\tbf{G}}$ (resp. ${\tbf{E}} \succ {\tbf{G}}$), if $E_r \ge  G_r \ \forall \ r$ 
(resp. if ${\tbf{E}} \succeq {\tbf{G}}$ and there exists some $r$ such that $E_r >  G_r$).
\end{definition}
%
%
%
\begin{lemma}
The SE operator of the coupled 
system maintains degradation in space, i.e., 
if  ${\tbf E} \succeq  {\tbf G}$, then $T_{\txt{c}}({\tbf E}) \succeq T_{\txt{c}}({\tbf{G}})$. 
This property is verified by $T_{\txt{u}}$ for a scalar error as well.
\label{lemma:spaceDegrad}
\end{lemma}
\begin{IEEEproof}
From \eqref{equ:effectivenoisecoupled} it is immediately seen that ${\tbf E} \succeq  {\tbf G}$ implies 
$\Sigma_c({\tbf E}) \geq \Sigma_c({\tbf G})\ \forall \ c$. Now, the SE operator \eqref{equ:stateevopcoupled} can be interpreted 
as an average over the spatial dimension of local MMSE's. The local MMSE's are the ones 
of $B$-dimensional AWGN channels with effective noise $\mathcal{N}(0,\Sigma_c^2/\log_2(B))$. The 
MMSE's are increasing functions of $\Sigma_c^2$: this is intuitively clear but we provide an explicit formula for the derivative below.
Thus $[T_{\txt{c}}( {\tbf E})]_r \geq [T_{\txt{c}}( {\tbf G})]_r \ \forall \ r$, which means $T_{\txt{c}}({\tbf E}) \succeq T_{\txt{c}}({\tbf{G}})$.

The derivative of the MMSE of the Gaussian channel with i.i.d noise $\mathcal{N}(0, \Sigma^2)$ can be computed as 
\begin{align}
 \frac{d \ \! {\rm mmse}(\Sigma)}{d(\Sigma^{-2})} = - 2 \mathbb{E}_{\bX, {\tbf Y}}\bigl[\|\bX - \mathbb{E}[\bX\vert {\tbf Y}]\|_2^2 {\rm Var}[\bX\vert {\tbf Y}]\bigl].
\end{align}
This formula is valid for vector distributions $p_0(\bx)$, and in particular, for our $B$-dimensional sections. It explicitly
confirms that $T_{\rm u}$ (resp. $[T_{\txt{c}}]_r$) is an 
increasing function of $\Sigma$ (resp. $\Sigma_c$). 
\end{IEEEproof}
\begin{figure}[!t]
\centering
\includegraphics[width=.5\textwidth]{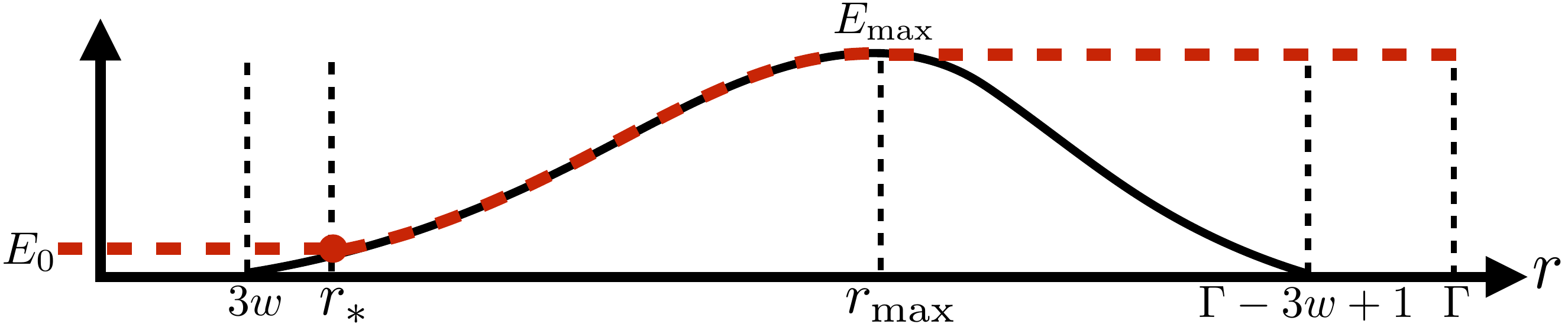}
\caption{A profile $\tbf{E}^{*}$ (solid) and its associated \emph{saturated} non 
decreasing profile $\tbf E$ (dashed). The former has a plateau at $0$ for all $r \le 3w$ and it increases until $r_{\rm max}$ where it reaches its maximum value $E_{{\rm max}}$. Then it decreases to $0$ at $\Gamma-3w+1$ and remains null after. The saturated profile starts with a plateau at $E_0$ for all $r \le r_*$, where $r_*$ is the position defined by 
$E^{*}_{r_*}=E_0$, and then matches $\tbf E^{*}$ for all $r\in\{r_*:r_{\rm max}\}$. It then saturates to $E_{{\rm max}}$ for all $r \ge r_{\rm max}$. By construction, ${\tbf E}$ is non decreasing and verifies 
${\tbf E} \succ {\tbf E}^{*}$.}
\label{fig:errorProfile}
\end{figure}
\begin{corollary}
\label{cor:timeDegrad}
The SE operator of the coupled 
system maintains degradation in time, 
i.e.,  $T_{\txt{c}}({\tbf E}^{(t)}) \preceq  {\tbf E}^{(t)}$, 
implies $T_{\txt{c}}({\tbf E}^{(t+1)}) \preceq  {\tbf E}^{(t+1)}$. 
Similarly $T_{\txt{c}}({\tbf E}^{(t)}) \succeq  {\tbf E}^{(t)}$ 
implies $T_{\txt{c}}({\tbf E}^{(t+1)}) \succeq  {\tbf E}^{(t+1)}$. 
Furthermore, the limiting error profile $ {\tbf E}^{(\infty)} \defeq T_{\txt{c}}^{(\infty)}({\tbf E}^{(0)})$
exists. 
This property is verified by $T_{\txt{u}}$ for the underlying system as well.
\end{corollary}
\begin{IEEEproof}
The degradation statements are a consequence of Lemma~\ref{lemma:spaceDegrad}. The existence 
of the limits follows from the monotonicity of the operator and boundedness of the MSE.
\end{IEEEproof}
\section{Proof of threshold saturation}\label{sec:proofsketch}
%
%
The pinning step together with the monotonicity properties of the coupled SE 
imply that the {\it fixed point} profile ${\tbf E}^{*}$ must adopt a shape similar to Fig.~\ref{fig:errorProfile} (note $E_{\rm max} \in [0,1]$). We associate to ${\tbf E}^{*}$ a {\it saturated profile} denoted $\tbf E$. Its construction is described in detail in Fig.~\ref{fig:errorProfile}. The saturated profile $\tbf E$ is strictly degraded with respect to ${\tbf E}^{*}$, thus $\tbf E$ serves as an upper bound in our proof.
\subsection{Coupled potential change evaluation by Taylor expansion}
\begin{definition}[Shift operator]
The \emph{shift operator} is defined pointwise as
$[\text{S}({\tbf E})]_1 \defeq E_0, \ [\text{S}({\tbf E})]_r \defeq  E_{r-1}$.
\end{definition}
\begin{lemma}\label{leminterp}
Let ${\tbf E}$ be a saturated profile. If $\hat {\tbf E} \defeq (1-\hat{t}) {\tbf E} + \hat{t}\text{S}({\tbf E})$ and $\Delta {E}_{r} \defeq {E}_{r} -{E}_{r-1}$, then for a proper $\hat{t}\in[0,1]$,
\begin{align*}
F_{\text{c}}(\text{S}({\tbf E})) &-F_{\text{c}}({\tbf E}) = \frac{1}{2} \sum_{r,r'=1}^{\Gamma} \Delta {E}_r \Delta {E}_{r'} \left[\frac{\partial^2 F_{\text{c}}}{\partial  E_{r}\partial  E_{r'}}\right]_{\hat {\tbf E}}.
\end{align*}
\label{lemma:Fdiff_quadraticForm}
\end{lemma}
\begin{IEEEproof}
$F_{\text{c}}(\text{S}({\tbf E})) - F_{\text{c}}({\tbf E})$ can be expressed as
\begin{align}
\frac{1}{2}\sum_{r,r'=1}^{\Gamma} \Delta {E}_{r} \Delta {E}_{r'}  \left[\frac{\partial^2 F_{\text{c}}}{\partial  E_{r}\partial  E_{r'}}\right]_{\hat {\tbf E}} - \sum_{r=1}^{\Gamma} \Delta {E}_{r}  \left[\frac{\partial F_{\text{c}}}{\partial  E_r}\right]_{{\tbf E}}. \label{eq:expFbsminusFb}
\end{align}
By saturation of $\tbf E$, $\Delta E_r=0 \ \forall \ r \in\mathcal{B}\defeq\{1:r_*\}\cup \{r_{\rm max}+1:\Gamma\}$. 
Moreover for $r\notin\mathcal{B}$, $E_r=[T_c(\tbf E)]_r$, and thus by Lemma~\ref{lemma:fixedpointSE_extPot} the potential 
derivative cancels at these positions. 
Hence the last sum in (\ref{eq:expFbsminusFb}) cancels.
\end{IEEEproof}
\begin{lemma}
The saturated profile ${\tbf E}$ is \emph{smooth}, i.e. $|\Delta {E}_{r} | < (g_* + \tilde{g}) /w$ where  
$\tilde{g}\defeq \max (1+ g_*, g_0 + 2g_*)$. Recall $g_0$ and $g_*$ are independent of $w$ and $\Gamma$.
\label{lemma:Evariesslowly}
\end{lemma}
\begin{IEEEproof}
$\Delta E_r=0$ for all $r \in\mathcal{B}$. By construction of $\{J_{r,c}\}$ and using Lipschitz continuity of $g_w$, we have for all $r \notin\mathcal{B}$ that $|\Delta {E}_{r}| = \big|\sum_{c} (J_{r,c} - J_{r-1,c}) \tilde E_c\big|/\Gamma < (g_* + \tilde{g})/w$.
\end{IEEEproof}
\begin{lemma}
\label{lemma:quadFormBounded}
The coupled potential verifies
\begin{align}  
\frac{1}{2}\Bigg|\sum_{r,r'=1}^{\Gamma} \Delta {E}_{r} \Delta {E}_{r'}\left[\frac{\partial^2 F_{\text{c}}}{\partial  E_{r}\partial  E_{r'}}\right]_{\hat{\tbf E}}\Bigg| < K/w, \label{eq:quadFormBounded}
\end{align}
where $K$ is independent of $w$ and $\Gamma$.
\end{lemma}

The proof uses Lemma~\ref{lemma:Evariesslowly} and the monotonicity of $E_r$. Bounding the Hessian 
is rather long but does not present major difficulties.
%
%
\subsection{Direct evaluation of the coupled potential change}
\begin{lemma}
Let ${\tbf E}$ be a saturated profile such that ${\tbf E}\succ {\tbf E}_0$. Then $ E_{\rm max} \notin \mathcal{V}_0$.
\label{lemma:outside_basin}
\end{lemma}
\begin{IEEEproof}
The non decreasing error profile and the assumption that $ {\tbf E} \succ {\tbf E}_0$ imply that $ E_{\rm max} > E_{0}$. Moreover,
$E_{\rm max} \le [T_\text{c}({\tbf E})]_{r_{\rm max}} \le T_\text{s}(E_{\rm max})$
where the first inequality follows from $\tbf E\succ \tbf E^*$ and the monotonicity of $T_\text{c}$, while the second comes from the variance symmetry at $r_{\rm max}$ and the fact that $\tbf E$
is non decreasing. Combining these with the 
monotonicity of $T_\text{s}$ gives: $T_{\text s}(E_{\rm max}) \ge  E_{\rm max} \Rightarrow T_{\text s}^{(\infty)}( E_{\rm max}) \ge  E_{\rm max}> E_0 \Rightarrow  E_{\rm max} \notin \mathcal {V}_0$.
\end{IEEEproof}
\begin{lemma}
Let ${\tbf E}$ be a saturated profile such that ${\tbf E}\succ {\tbf E}_0$. Then $F_{\text{c}}(\text{S}({\tbf E})) - F_{\text{c}}({\tbf E}) \leq -\Delta F_{\text{u}}.$
\label{lemma:diffShited_directEval}
\end{lemma}
\begin{IEEEproof}
The contribution of the change in the energy term is a perfect telescoping sum:
\begin{align}  
U_{\text{c}}(\text{S}({\tbf E})) - U_{\text{c}}({\tbf E}) = U_{\text{u}}(E_0) - U_{\text{u}}(E_{\rm max}). \label{eq:DeltaU}
\end{align}
We now deal with the contribution of the change in the entropy term. We first notice that, using the variance symmetry, $\Sigma_c({\tbf E}) = \Sigma_{c+1}(\text{S}({\tbf E})) \ \forall \ c \in \{2w + 1:\Gamma-2w-1\}$, which makes the sum telescoping in this set. Thus
\begin{align}  
&S_{\text{c}}({\tbf E})-S_{\text{c}}(\text{S}({\tbf E}))= S_{\text{u}}( \Sigma_{\Gamma-2w}({\tbf E}))-S_{\text{u}}( \Sigma_{2w+1}(\text{S}{(\tbf E)}))  \nonumber \\
&- \sum_{c\in \mathcal{S}} [S_{\text{u}}( \Sigma_{c}(\text{S}{(\tbf E)})) - S_{\text{u}}( \Sigma_{c}({\tbf E}))], \label{eq:diffSB}
\end{align}
where $\mathcal{S}\defeq\{1:2w\}\cup\{\Gamma-2w+1:\Gamma\}$. By saturation, $\tbf E$ possesses the following property: $[\text{S}( {\tbf E})]_r = [ {\tbf E}]_r \ \forall \ r \in \{1:r_*\}\cup\{r_{\rm max}+1:\Gamma\} \Rightarrow  \Sigma_c(\text{S}( {\tbf E})) =  \Sigma_c( {\tbf E}) \ \forall \ c \in \mathcal{S}$ and thus the sum in (\ref{eq:diffSB}) cancels. Furthermore, one can show, using the saturation of $\tbf E$ and the variance symmetry, that $\Sigma_{2w+1}(\text{S}({\tbf E})) = \Sigma( E_{0})$. Using the same arguments, in addition to the fact that $\tbf E\succ \tbf{E}_0\Rightarrow r_{\rm max} \le \Gamma-3w$, we obtain $\Sigma_{\Gamma-2w}({\tbf E}) = \Sigma( E_{\rm max})$.
Hence, (\ref{eq:diffSB}) yields the following:
\begin{align}  
S_{\text{c}}({\tbf E}) - S_{\text{c}}(\text{S}({\tbf E})) = S_{\text{u}}(\Sigma( E_{\rm max})) - S_{\text{u}}(\Sigma( E_{0})).
\label{eq:DeltaS}
\end{align} 
Combining (\ref{eq:DeltaU}) with (\ref{eq:DeltaS}) and using Lemma~\ref{lemma:outside_basin} gives
\begin{align*}  
F_{\text{c}}(\text{S}({\tbf E})) - F_{\text{c}}({\tbf E}) = -F_{\text{u}}( E_{\rm max}) + F_{\text{u}}({E_0}) \leq -\Delta F_{\text{u}}.
\end{align*}
\end{IEEEproof}
\begin{thm}
Assume a spatially coupled SS code ensemble is used for communication through 
an AWGNC. Fix $R<R_{\txt{pot}}$, $w>K/\Delta F_{\text{u}}$ ($K$ is independent of $w$ and $\Gamma$)
and $\Gamma> 8w$ (such that the code is well defined). Then any fixed point error profile ${\tbf{E}^*}$ of the coupled SE satisfies ${\tbf{E}}^* \prec {\tbf{E}}_0$.
\label{th:mainTheorem}
\end{thm}
\begin{IEEEproof}
Assume that, under these hypotheses, there exists 
a saturated profile $\tbf E$ associated to $\tbf E^*$ such 
that $\tbf E\succ {\tbf{E}}_0$. By Lemma~\ref{lemma:diffShited_directEval} and 
the positivity of $\Delta F_{\text{u}}$ as $R<R_{\txt{pot}}$, 
we have $|F_{\text{c}}({\tbf E}) - F_{\text{c}}(\text{S}({\tbf E}))| \ge \Delta F_{\text{u}}$. 
Therefore, by Lemmas~\ref{lemma:Fdiff_quadraticForm} and \ref{lemma:quadFormBounded} we get $K/w > \Delta F_{\text{u}} \Rightarrow w < K/\Delta F_{\text{u}}$, a contradiction. 
Hence, ${\tbf{E}} \preceq {\tbf{E}}_0$. Since $\tbf E \succ {\tbf E}^*$ we have ${\tbf E}^* \prec {\tbf{E}}_0$.
\end{IEEEproof}
\begin{corollary}\label{cor:maincorollary}
By first 
taking $\Gamma \to \infty$ and then $w\to\infty$, the AMP threshold of the coupled 
ensemble  satisfies $R_{\text c}\geq R_{\text{pot}}$.
\end{corollary}

This result follows from Theorem~\ref{th:mainTheorem} and Definition~\ref{def:AMPcoupled}. It says that the AMP threshold for the coupled codes saturates the potential threshold. To prove  
it cannot surpass it requires a separate treatment. For our purposes this is not really needed because 
we have necessarily $R_{\text c} < C$ and we know (Sec.~\ref{sec:larg_B}) that $\lim_{B\to \infty}R_{\text{pot}}=C$. Thus $\lim_{B\to \infty} R_{\text c} = C$.

%
%
\section*{Acknowledgments}
J.B and M.D acknowledge funding from the Swiss National Science Foundation grant num. 200021-156672. 
We thank Marc Vuffray and Rüdiger Urbanke for discussions.
\ifCLASSOPTIONcaptionsoff
\fi
\bibliographystyle{IEEEtran}
\bibliography{IEEEabrv,bibliography}
\end{document}